\documentstyle [preprint,aps,eqsecnum]{revtex}

\def\rate{\langle R \rangle}
\def\dk{{d^3 k \over (2 \pi)^3}}

\def\omgtk{\omega_{\rm t} ({\bf k})}
\def\omglk{\omega_{\rm l} ({\bf k})}
\def\h{{\rm h}}
\def\t{{\rm t}}
\def\l{{\rm l}}
\def\s{{\rm s}}
\def\e{{\rm e}}

\begin{document}

\preprint{PURD-TH-96-07}

\title{Transition rate for process involving
particles with high momentum in a plasma
and infrared physics for QED plasma}

\author{Chengxing Zhai}

\address{Department of Physics,
	 Purdue University,
	 West Lafayette, IN 47907}

\maketitle

\begin{abstract}

We derive a formula for computing the transition rate for a process
involving particles with momentum much higher than the temperature and
chemical potentials in a plasma by using an effective field theory
approach. We apply it to collision of charged particles with hard
momentum inside a QED plasma.  The Debye screening effect and the
damping of a charged particle moving in QED plasma are studied.  Using
the Bloch-Nordsieck resummation, the infrared divergences due to the
absence of magnetic screening for QED plasma are shown not to appear in
physically measurable rates.  The soft plasmon absorption and emission
for charged particles are discussed.

\end{abstract}

\newpage

\section{Introduction}
Effective field theory has been proved to be powerful in
dealing with different energy scales. Consider a process
that contains particles with hard momentum (hard momentum
means one of the component of the four momentum is larger
than the temperature and the chemical potentials.)
happening inside of a plasma. There are two relevant energy
scales: the hard momentum scale and the soft momentum scale
which is the thermal scale, {\it i.e.} order of the temperature
$T$ or the chemical potential.
It is convenient to first ``integrate out'' the hard field
degree of freedom with the momentum scale much higher than
the temperature and the chemical potentials and end up with
an effective action for the soft field degree of freedom.
Taking the thermal average of the soft field degree of freedom
yields the transition rate for the process.
In this way, we can deal with the hard momentum scale
and the thermal scale separately.
Applying this formula to the hard process involving charged
particle inside a QED plasma automatically incorporates the Debye
screening effect and enables us to study
the infrared physics for QED plasma due to the
exchange of soft photon between the charged particles and the plasma,
such as the damping of a charged particle moving in QED plasma
as well as the emission and absorption of soft plasmon.

It is well known that the infrared divergences coming from
the radiative correction due to the virtue soft photon are canceled by
the real soft photon emission\cite{Bloch&Nordsieck,Yennie}.
When the system is heated to a finite temperature, it is even more
sensitive to the infrared region for the photons because of the
bosonic distribution factor $n(\omega) \sim T/\omega$.
It is interesting then
to examine the cancellation of the infrared divergences for QED
at finite temperature where a large population of thermal soft photons
due to the Bose distribution enhances both the emission and the
absorption of the soft photons for the charged particles.
In the literature, there has been study of this problem for the
free thermal photon background\cite{Weldon1,Landshoff&Taylor}.
Our formula enables us to provide an analysis of these emission
and absorption processes for the interacting
photons inside QED plasma. The interaction between photons and
the plasma brings the Debye screening mechanism to shield
the electric force but not the magnetic force. Therefore,
potential infrared problem could occur due to the exchange of
soft magnetic photons. In fact, there have been many discussions
on the infrared divergence for the damping rate of a charged particle
moving in QED plasma\cite{Lebedev&Smilga&Baier}.
As discussed later in section \ref{damping_subsec},
this arises because the unscreened magnetic force
does not allow the charged particle moving with a definite momentum.
For a physically measurable damping rate, this infrared divergence is cut
off by the momentum resolution of our experiment.
In general, we show that the infrared divergences
due to the unscreened magnetic force cancels out by using
the Bloch-Nordsieck resummation. We also discuss
the correction to the transition rate due to
the soft plasmon emission and absorption.
At the massless limit of the plasmon, we reproduce the previous
result\cite{Weldon2}.

The organization of this paper is as following.
In next section, we develop first a general
formula for calculating the transition rate for
a hard process in a thermal background.
In section III, our formula is applied to process inside QED
plasma to study the Debye screening effect, the damping
of a charged particle moving inside QED plasma, and
infrared physics of QED plasma. Conclusions are drawn in section IV.

\section{Thermal average of transition rate}

In this section, we derive a general formula for
computing the transition rate for a process involving
hard momentum particles with the appearance of a
relatively low (compared with the hard momentum scale) temperature $T$
thermal background. The interaction between the thermal
background and the hard particles will be considered.

Let us denote the fields with hard momentum by $\phi_\h$
and the fields with soft momentum by $\phi_\s$.
Since, we consider the case where the temperature is
much lower than the hard momentum, for the hard momentum
fields, it is valid to treat them as at zero temperature.
The thermal averaged transition rates reads
\begin{equation}
    \rate = {1 \over T_\t} {1 \over Z} \sum_{m,n} |\langle f, m |
	i, n \rangle |^2 e^{-\beta ( E_n - \sum_i \mu_i Q_i(n))}
\label{rate_def}
\end{equation}
where $T_t$ is the total time for the interaction to take place and
$i, f$ label the initial and final states of the
hard momentum particles respectively. $m, n$ represent
the energy eigenstates generated by the soft momentum fields $\phi_\s$. 
A thermal distribution factor has been included with $\beta = 1/T$,
$\mu_i$ are the chemical potentials corresponding to conserved
charges $Q_i$. $E_n$ and $Q_i(n)$ are the eigenvalues of the Hamiltonian
and the charge $Q_i$ for state $| n \rangle$.
$Z$ is the partition function for the soft momentum fields
\begin{equation}
    Z = Tr \left (e^{-\beta(H - \sum_i \mu_i Q_i)} \right )
	= \sum_n e^{-\beta (E_n - \sum_i \mu_i Q_i (n))} \,.
\end{equation}
To be more specific, we shall consider $n_i$ initial
and $n_f$ final hard momentum particles are involved.
Let us use $p'_f$ with $f=1,2,\cdots,n_f$ and $p_i$ with $i=1,2,\cdots,n_i$
denote the momentum of the final and initial particles.
Therefore,
\begin{equation}
    \rate = {1 \over T_\t} {1 \over V^{n_i}} {1 \over Z}
	\prod_{f=1}^{n_f} \int_{\Omega_f} \! {d^3 p'_f \over (2 \pi)^3}
	\sum_{n,m}
	|\langle m, p'_1, p'_2, \cdots, p'_{n_f} | p_1,p_2,\cdots,p_{n_i},
	n \rangle |^2 e^{-\beta (E_n - \sum_i \mu_i Q_i (n))}
\label{thermo_rate}
\end{equation}
where $\Omega_f$ are the phase space volume within which the detectors
are hunting for the final products.
$V$ is the total volume of the spatial box in which the wave function
of the initial and final states are normalized.
Expression (\ref{thermo_rate}) has following structure:
matrix element $\langle m, p'_1, p'_2, \cdots, p'_{n_f} |
p_1,p_2,\cdots,p_{n_i}, n \rangle$ involves time evolution
from far past to far future while its complex conjugate evolves
from far future to far past; the thermal distribution factor
$\exp \{-\beta(E_n - \sum_i \mu_i Q_i (n))\}$ causes an evolution
in imaginary time by amount $i\beta$. A real time formalism
for the thermal field theory\cite{Rivers&Landsman&Weert}
can conveniently incorporate this structure so that we
can express it in terms of functional integrals defined along
paths in complex time plane.
Slicing the time evolutions into pieces by inserting complete
sets of states as we usually do to derive the path integral formalism
and using the reduction formula to write the matrix element
in terms of Green's function, Eq.~(\ref{thermo_rate}) may be expressed as
\begin{eqnarray}
    \rate &=& {1 \over T_\t} {1 \over V^{n_i}}
	\prod_{f=1}^{n_f} \int_{\Omega_f} {d^3 p_f \over (2 \pi)^3}
	\int_P {\cal D} \phi_\s
\nonumber\\
        && \times \int_{P_1} {\cal D} \phi_\h
	\prod_{f=1}^{n_f} \int d^3{\bf x}_f g_f^*(p'_f,x_f)
	{1 \over i} {\stackrel \leftrightarrow \partial}_{x^0_f}
	\phi_\h (x_f) \;
	\prod_{i=1}^{n_i} \int d^3 {\bf x}_i f_i (p_i,x_i)
	{1 \over i} {\stackrel \leftrightarrow \partial}_{x^0_i}
	\phi_\h (x_i)
\nonumber\\
	&& \times \int_{P_2} {\cal D} \phi_\h
	\prod_{i=1}^{n_i} \int d^3 {\bf x}'_i f^*_i (p_i,x'_i)
	\left (- {1 \over i}
	{\stackrel \leftrightarrow \partial}_{x^{\prime 0}_i} \right )
	\phi^*_\h (x'_i)\;
        \prod_{f=1}^{n_f} \int d^3{\bf x}'_f g_f (p'_f,x'_f)
	\left (- {1 \over i}
	{\stackrel \leftrightarrow \partial}_{x^{\prime 0}_f} \right )
	\phi^*_\h (x'_f)
\nonumber\\
	&& \times \exp \left \{ i S_{P}[\phi_\s]
	+ i S_{P_1} [\phi_\h,\phi_\s] + i S_{P_2} [\phi_\h,\phi_\s]
	\right \} \,.
\end{eqnarray}
Here the notations need to be explained.
$f_i(p_i,x_i)$ and $g_f(p_f,x_f)$ are the wave functions for the
initial and final single particle states respectively.
The path integrals are along contours in complex time denoted as $\tau$
plane: contour $P_1$ starts from $\tau{=}{-}\infty$ to
$\tau{=}{+}\infty{-}i\epsilon$; contour $P_2$ starts from
$\tau{=}{+}\infty{-}i\epsilon$ to $\tau{=}{-}\infty{-}2 i\epsilon$;
contour $P$ includes $P_1$, $P_2$, and another piece from
$\tau{=}{-}\infty{-}2 i\epsilon$ to $\tau{=}{-}\infty{-}i \beta$.
Boundary conditions are $\phi_\s (\tau{=}{-}\infty,{\bf x}){=}
\pm \phi_\s (\tau{=}{-}\infty{-}i\beta, {\bf x})$
for the soft bosonic (+) and fermionic ($-$) momentum fields
while $\phi_\h (\tau{=}\pm \infty, {\bf x}) \to 0$ for hard momentum fields.
Actions $S_{P} [\phi_\s]$, $S_{P_1} [\phi_\h,\phi_\s]$, and
$S_{P_2} [\phi_\h,\phi_\s]$ are defined with the
time integral being the contour integral in the complex time $\tau$
plane along $P,P_1,P_2$ respectively. $S_{P} [\phi_\s]$
represents the part of the action contains only $\phi_\s$
while $S_{P_1} [\phi_\h,\phi_\s]$ and $S_{P_2} [\phi_\h,\phi_\s]$
contains both the part containing only $\phi_\h$ and the interaction
terms between $\phi_\h$ and $\phi_\s$. The time component of vectors
$x_i,x_f$ belong to path $P_1$ while the time component of vectors
$x'_i,x'_f$ belong to path $P_2$.
This type of path integral is used usually in the real time formalism
for thermal field theory\cite{Rivers&Landsman&Weert}.

From an effective field theory point of view, we can first
integrate out the hard momentum degree of freedom.
The formula above may be rewritten as
\begin{eqnarray}
    \rate &=& {1 \over T_\t} {1 \over V^{n_i}}
        \prod_{f=1}^{n_f} \int_{\Omega_f} {d^3 p_i \over (2 \pi)^3}
        \int_P {\cal D} \phi_\s e^{i S_{P}[\phi_\s] }
        \langle p'_1, p'_2, \cdots, p'_{n_f}
	| p_1,p_2,\cdots,p_{n_i} \rangle_{\phi_\s, P_1}
\nonumber\\
	&& \times \langle  p_1,p_2,\cdots,p_{n_i}
         | p'_1, p'_2, \cdots, p'_{n_f} \rangle_{\phi_\s, P_2}
\end{eqnarray}
or more compactly
\begin{equation}
    \rate = {1 \over T_t} \int_P {\cal D} \phi_\s
	\exp \left \{ i S_{P}[\phi_\s] \right \}
	\langle f | i \rangle_{\phi_\s,P_1}
	\langle i | f \rangle_{\phi_\s,P_2} \,,
\label{contour}
\end{equation}
where $\langle f | i \rangle_{\phi_\s,P_1}$
represents the matrix element between the hard momentum states
$| i \rangle$ and $\langle f |$ with the background soft momentum
field $\phi_\s$ appearing. Subscripts $P_1, P_2$ denote the time argument in
field $\phi_\s$ belonging to contours $P_1, P_2$ respectively.
We note that the thermal effect has been taken into account
by the ``strange'' path integral that we use and 
the chemical potential for the conserved charges are incorporated
by including them in the action $S_{P}[\phi_\s]$ as the
Lagrangian multipliers.
It is convenient to introduce source terms for field $\phi_\s$
with time argument belonging to paths $P_1$ and $P_2$ respectively
and define following generating functional
\begin{equation}
    e^{i W[j_\s, j'_\s]} \equiv \int_P {\cal D} \phi_\s
	\exp \left \{ i S_{P}[\phi_\s]
	{+}i \!\int_{P_1}\!\! d\tau \!\int\! d{\bf x}
	j_\s (\tau,{\bf x}) \phi_\s (\tau,{\bf x})
	{+}i \!\int_{P_2}\!\! d\tau' \!\int\! d{\bf x}'
	j'_\s (\tau',{\bf x}') \phi^*_\s (\tau',{\bf x}')\right \} \,.
\label{W_def}
\end{equation}
With this,
\begin{equation}
    \rate = {1 \over T_\t} \left . {\cal A}_{fi}
	\left[ {1 \over i} {\delta \over \delta j_\s}\right ]
	{\cal A}^*_{if} \left [{1 \over i} {\delta \over
	\delta j'_\s} \right ]
	e^{i W[j_\s,j'_\s]} \right |_{j_\s=0,j'_\s=0} \,,
\label{amp}
\end{equation}
where the functional ${\cal A}_{fi}[\phi_\s]$ is defined as
\begin{equation}
    {\cal A}_{fi} [\phi_\s] \equiv \langle f | i \rangle_{\phi_\s} \,.
\end{equation}
When the argument for ${\cal A}_{fi}$ is a functional
derivative operator, the power series expansion of the functional
is understood.
We note that the path integral along contour $P$ is simply another
way to write the thermal average. Generating functional~(\ref{W_def})
may be written as a thermal average in an operator form:
\begin{equation}
    e^{i W[j_\s,j'_\s]} {=}
	{1 \over Z} Tr\!\left [e^{-\beta(H{-}\sum_i \mu_i Q_i)}
	T_{+} \!\left (e^{i \!\int_{-\infty}^{\infty}\!\! dt \!\int d{\bf x}
	j_\s(t,{\bf x}) \phi_\s(t,{\bf x})} \right)
	T_{-} \!\left (e^{{-} i \!\int_{-\infty}^{\infty}\!\! dt'
	\!\int d{\bf x}' j_\s(t',{\bf x}')
	\phi_\s^\dagger({\bf x}',t')} \right ) \right ]
\end{equation}
where $T_{+}$ means time-ordered product according to $t$
and $T_{-}$ means anti-time-ordered product according to $t'$.
Here the Hamiltonian $H$ and the charges $Q_i$ are operators
for the degree of freedom corresponding to field operator $\phi_\s$.

The formula above enables us to deal with the two scales involved
in the process separately. The characteristic energy scale for
amplitude $\langle f | i \rangle_{\phi_\s,P_1}$ is the hard momentum
scale. We can then treat $\phi_\s$ as a slowly variant background
field which enables us to use, for example, the semi-classical method.
After we completed the hard momentum scale, we only need to deal with
the soft momentum scale physics left.

We note that it does not require the momentum transfer
to be big to get the formula above. All we need is that the
particles involved has high momentum so that the thermal bath
does not contain a noticeable amount of these hard momentum
particles which we want to scatter and detect.
This formula is also valid for the case where we just formed
a plasma which is weakly coupled to another kind of particle,
for example, the neutrinos.
The plasma does not have enough time to produce the thermo-neutrinos,
so that we can treat the neutrino fields the same way as $\phi_\h$
without requiring neutrinos to carry only hard momentum
because the neutrino field is effectively at zero temperature.
Practically, we can do a perturbation in
the weak coupling constant to evaluate $\langle f|i \rangle_{\phi_\s}$
for using this formulation.

A concrete formulation for computing non-relativistic reaction rate
has been well developed in reference\cite{Brown&Sawyer}.

\section{Process for hard charged particles in QED plasma}

We now apply the formulation in previous section to
study explicitly a collision involving hard charged particles
with the appearance of QED plasma.
Let the process start with $n_i$ initial particles and end up with
$n_f$ final particles plus soft photons.
We assume that the initial particles have the four-momentum
$p_i$, with $i=1,2,\cdots,n_f$ and we detect the final particles
with momentum $p'_f$ with $f=1,2,\cdots,n_f$ but left the soft
photons undetected. In view of the formula in the previous section,
$\phi_\s$ now represents the thermal photon and
thermal electron fields while $\phi_\h$ represents
the hard charged particles. 
Integrating out the degree of freedom corresponding to $\phi_\h$
to get an effective functional integrand for $\phi_\s$ means
to get the zero temperature transition rate for hard charged particles
with the appearance of the thermal QED background field.
Expression~(\ref{contour}) for transition rate reads
\begin{eqnarray}
    \rate &=& {1 \over T_\t} {1 \over V^{n_i}}
	\int_P {\cal D} A {\cal D} \bar \psi {\cal D} \psi
	\exp \left \{i S^{\rm eff}_P [A, \bar \psi, \psi] \right \}
\nonumber\\
        && \times \prod_{f=1}^{n_f} \int dy_f \phi^{0\,*}_\h (p'_f, y_f)
        (\partial_{y_f}^2{+}m^2)
        \prod_{i=1}^{n_i} \int dx_i \phi^0_\h (p_i, x_i)
        (\partial_{x_i}^2{+}m^2)
\nonumber\\
        && \times \prod_{i=1}^{n_i} \int dx'_i \phi^{0\,*}_\h (p_i, x'_i)
        (\partial_{x'_i}^2{+}m^2) 
        \prod_{f=1}^{n_f} \int dy'_f \phi^0_\h (p'_f, y'_f)
        (\partial_{y'_f}^2{+}m^2)
\nonumber\\
	&& \times
	G[x_1,x_2,\cdots,x_{n_i},y_1,y_2,\cdots,y_{n_f}; A_{P_1}]
	G^*[x'_1,x'_2,\cdots,x'_{n_i},y'_1,y'_2,\cdots,y'_{n_f}; A_{P_2}] \,.
\end{eqnarray}
Here $G[x_1,x_2,\cdots,x_{n_i},y_1,y_2,\cdots,y_{n_f}; A]$
is the Green's function for the
hard particle fields with background photon field $A$.
$A_{P_1}$ and $A_{P_2}$ denote
the photon fields with the time argument being on
the contours $P_1$ and $P_2$ respectively.
$\phi^0_\h(p,x)$ is the wave function for the charged particle
moving freely.
The integration of the hard fields has two effects:
1) it gives the Green's function $G[x_1,x_2,\cdots,x_{n_i},
y_1,y_2,\cdots,y_{n_f};A]$;
2) it gives radiative corrections to the action for
the photon and electron fields, i.e. we have an effective
action. Except the effect of renormalization, these corrections are
usually small because the high energy mode
decouples\cite{Applequist&Carrazone}.

Since $G[x_1,x_2,\cdots,x_{n_i},y_1,y_2,\cdots,y_{n_f};A]$
contains only photon field,
it is convenient to further integrate out the electron degree of freedom.
Introducing source terms for the photon fields on paths
$P_1$ and $P_2$, formally, we arrive at
\begin{eqnarray}
    \rate &=& {1 \over T_\t} {1 \over V^{n_i}}
        \prod_{f=1}^{n_f} \int dy_f \phi^{0\,*}_\h (p'_f, y_f)
        (\partial_{y_f}^2{+}m^2)
        \prod_{i=1}^{n_i} \int dx_i \phi^0_\h (p_i, x_i)
        (\partial_{x_i}^2{+}m^2)
\nonumber\\
        && \times \prod_{i=1}^{n_i} \int dx'_i \phi^{0\,*}_\h (p_i, x'_i)
        (\partial_{x'_i}^2{+}m^2) 
        \prod_{f=1}^{n_f} \int dy'_f \phi^0_\h (p'_f, y'_f)
        (\partial_{y'_f}^2{+}m^2)
\nonumber\\
	&& \times G \left [x_1,x_2,\cdots,x_{n_i},y_1,y_2,\cdots,y_{n_f};
        {1 \over i} {\delta \over \delta J} \right ]
\nonumber\\
	&& \times
	G^*\left [x'_1,x'_2,\cdots,x'_{n_i},y'_1,y'_2,\cdots,y'_{n_f};
        {1 \over i} {\delta \over \delta J'} \right ]
	\left . e^{ i W [J,J'] } \right |_{J{=}0,J'{=}0}  \,,
\end{eqnarray}
where the generating functional $W[J,J']$ is defined by
\begin{eqnarray}
    \exp \{ i W[J,J']\} &\equiv&
	\int_P {\cal D} A {\cal D} \bar \psi {\cal D} \psi
	\exp \left \{i S^{\rm eff}_P [A, \bar \psi, \psi]
	+ i \int_{P_1} J \cdot A
	+ i \int_{P_2} J' \cdot A \right \}
\nonumber\\
	&=& \left \langle
	T_+ \left (e^{i \int dt \int d{\bf x} J (t,{\bf x})
	\cdot A (t,{\bf x})} \right )
	T_- \left (e^{- i \int dt' \int d{\bf x}' J' (t',{\bf x}')
	\cdot A (t',{\bf x}')} \right ) \right \rangle \,,
\end{eqnarray}
where we have used $\langle \cdots \rangle$ to denote the thermal
average.
Power series expansion of $W[J,J']$ can be written as a quadratic term
plus higher power order terms in source fields $J,J'$:
\begin{eqnarray}
    W [J,J'] &=& - {1 \over 2} \int d{\bf x}_1 \int d{\bf x}_2
	\Big [ \int dt_1 \int dt_2
	J_\mu (t_1,{\bf x}_1) J_\nu (t_2,{\bf x}_2)
	G^{(+)}_{\mu\nu} (t_1,{\bf x}_1;t_2,{\bf x}_2)
\nonumber\\
	&& + \int dt'_1 \int dt'_2
	J'_\mu (t'_1,{\bf x}_1) J'_\nu (t'_2,{\bf x}_2)
	G^{(-)}_{\mu\nu} (t'_1,{\bf x}_1;t'_2,{\bf x}_2)
\nonumber\\
	&& - 2 \int dt_1 \int dt'_2
	J_\mu (t_1,{\bf x}_1) J'_\nu (t'_2,{\bf x}_2)
	{\cal G}_{\mu\nu} (t_1,{\bf x}_1;t'_2,{\bf x}_2) \Big ]
	+ \cdots
\label{quadratic}
\end{eqnarray}
where
\begin{eqnarray}
    G^{(+,-)}_{\mu\nu} (t_1,{\bf x}_1;t_2,{\bf x}_2)
	&\equiv& \langle T_{(+,-)} \left ( A_\mu(t_1,{\bf x}_1)
	A_\nu (t_2,{\bf x}_2) \right ) \rangle \,,
\nonumber\\
    {\cal G}_{\mu\nu} (t_1,{\bf x}_1;t'_2,{\bf x}_2)
	&\equiv& \langle A_\mu(t_1,{\bf x}_1)
 	A_\nu (t'_2,{\bf x}_2) \rangle \,.
\end{eqnarray}
Again, we use $+,-$ to denote the time-ordered and anti-time-ordered
product.
We have following relations
\begin{equation}
    G^{(+,-)}_{\mu\nu} (t_1,{\bf x}_1;t_2,{\bf x}_2)
	= \int \dk e^{i {\bf k} \cdot ({\bf x}_1 - {\bf x}_2)}
	{\cal G}_{\mu\nu} (t_{(>,<)}, t_{(<,>)}; {\bf k})
\end{equation}
where $t_> = {\rm max}\{t_1,t_2\}$ and $t_< = {\rm min} \{t_1,t_2\}$
and ${\cal G}_{\mu\nu} (t,t';{\bf k})$ is defined by
\begin{equation}
    {\cal G}_{\mu\nu} (t,{\bf x};t',{\bf x}')
	= \int \dk e^{i {\bf k} \cdot ({\bf x} - {\bf x}')}
	{\cal G}_{\mu\nu} (t,t';{\bf k}) \,.
\end{equation}
The Green's function ${\cal G}_{\mu\nu}$ may be expressed
in terms of the spectral density function:
\begin{equation}
    {\cal G}_{\mu\nu} (t, t', {\bf k})
	= \int_{-\infty}^\infty {d \omega \over 2 \pi i} \int \dk n(\omega)
	\rho_{\mu\nu} (\omega,{\bf k}) e^{-i \omega (t - t')} \,,
\end{equation}
where $\rho_{\mu\nu}(\omega,{\bf k})$ is the usual
spectral density function and $n(\omega)$ is the Bose distribution factor
\begin{equation}
    n(\omega) = {1 \over e^{\beta \omega} - 1} \,.
\end{equation}

We shall focus on two aspects due to plasma. The first
is for the momentum transfer is comparable with the
Debye screening length, we show that the Debye screening
effect can be properly taken into account in our formalism.
The second case that we shall study is the modification to
the transition rate due to the soft photon.

\subsection{Debye screening}

For the reason of simplicity, we consider the collision of
two spin ${1 \over 2}$ particles and ignore the soft photon
emission and absorption which will be considered later.
We have particle 1 with charge $q_1$ and 2 with charge $q_2$
moving with momentum $p_1$ and $p_2$ initially, colliding with each other
via EM interaction, and separating with momentum $p'_1$ and $p'_2$
finally.  When the momentum transfer is comparable to the Debye length for
QED plasma, we shall be able to probe the Debye screening effect.
We do a perturbation theory in QED coupling constant.
At the leading order,
\begin{eqnarray}
    \langle p'_1, p_2' | p_1, p_2 \rangle_{A, P_1}
	&=& - q_1 q_2 \int d x_1 e^{-i (p_1{-}p'_1) \cdot x_1}
	A_\mu (x_1) \, \Phi^{1*}_0 (p'_1) \gamma^\mu \Phi^1_0 (p_1)
\nonumber\\
	&& \qquad \qquad \times \int d x_2 e^{-i (p_2{-}p'_2) \cdot x_2}
	A_\nu (x_2) \, \Phi^{2*}_0 (p'_2) \gamma^\nu \Phi^2_0 (p_2)
\end{eqnarray}
where $\Phi_0^{1,2}$ are the wave functions of the particles 1 and 2
moving freely with specific momentum respectively.
Expression~(\ref{amp}) for the rate reads
\begin{eqnarray}
    \rate &=&{1 \over T_\t}{1 \over V^2}
	\int_{\Omega_1} {d^3 p'_1 \over (2 \pi)^3}
	\int_{\Omega_2} {d^3 p'_2 \over (2 \pi)^3}
	\int \! d x_1 \! \int \! dx_2 \!
	\int \! d x'_1 \! \int \! d x'_2 \, 
	e^{i (p_1{-}p'_1) (x_1{-}x'_1)
	+ i (p_2{-}p'_2) (x'_2{-}x_2)}
\nonumber\\
	&& \times
	q_1^2 \, q_2^2 \,
	{\bar \Phi}^1_0 (p'_1) \gamma^\mu \Phi^1_0 (p_1) \,
	{\bar \Phi}^1_0 (p_1) \gamma^\nu \Phi^1_0 (p'_1) \,
	{\bar\Phi}^2_0 (p'_2) \gamma^{\lambda} \Phi^2_0 (p_2) \,
	{\bar\Phi}^2_0 (p_2) \gamma^{\sigma} \Phi^2_0 (p'_2)
\nonumber\\
	&& \times \left . {\delta \over \delta J_{\mu}(x_1)}
	{\delta \over \delta J_{\lambda}(x_2)}
	{\delta \over \delta J'_{\nu}(x'_1)}
	{\delta \over \delta J'_{\sigma}(x'_2)}
	e^{i W[J,J']} \right |_{J{=}0,J'{=}0} \,,
\end{eqnarray}
where $\Omega_{1,2}$ are phase space volumes for particles 1 and 2
and the usual notation
$\bar \Phi \equiv \Phi^\dagger \gamma^0$ has been used.
Since the four-point connected Green's function for photon fields is
of higher order, we shall neglect it.
Upon using Eq.~(\ref{quadratic}),
\begin{eqnarray}
    \rate &=&{1 \over T_\t}{1 \over V^2}
	\int_{\Omega_1} {d^3 p'_1 \over (2 \pi)^3}
	\int_{\Omega_2} {d^3 p'_2 \over (2 \pi)^3}
	\int \! d x_1 \! \int \! dx_2 \!
	\int \! d x'_1 \! \int \! d x'_2 \, 
	e^{i (p_1{-}p'_1) (x_1{-}x'_1)
	+ i (p_2{-}p'_2) (x'_2{-}x_2)}
\nonumber\\
	&& \times
	q_1^2 \, q_2^2 \,
	{\bar \Phi}^1_0 (p'_1) \gamma^\mu \Phi^1_0 (p_1) \,
	{\bar \Phi}^1_0 (p_1) \gamma^\nu \Phi^1_0 (p'_1) \,
	{\bar\Phi}^2_0 (p'_2) \gamma^{\lambda} \Phi^2_0 (p_2) \,
	{\bar\Phi}^2_0 (p_2) \gamma^{\sigma} \Phi^2_0 (p'_2)
\nonumber\\
	&& \times \left [G^{(+)}_{\mu\lambda}(x_1,x_2)
	G^{(-)}_{\nu\sigma}(x'_1,x'_2)
	+ {\cal G}_{\mu\nu}(x_1,x'_1)
	{\cal G}_{\lambda\sigma}(x_2,x'_2)
	+ {\cal G}_{\mu\sigma}(x_1,x'_2)
	{\cal G}_{\lambda\nu}(x_2,x'_1) \right ] \,.
\end{eqnarray}
The second term in the last square bracket corresponds to the
process with particle 1 and 2 scattering with the plasma
independently. The third term is an enhancing term for the special case
with particle 1 and 2 scattering with the plasma
independently but emitting the photon with the same four-momenta.
We shall consider only the part in the transition rate which
involves the interaction between particle 1 and 2. This corresponds
to the first term. Therefore,
\begin{eqnarray}
    \rate_{\rm int} &=&{1 \over T_\t}{1 \over V^2}
	\int_{\Omega_1} {d^3 p'_1 \over (2 \pi)^3}
	\int_{\Omega_2} {d^3 p'_2 \over (2 \pi)^3} \,
	q_1^2 \, q_2^2 \,
	{\bar \Phi}^1_0 (p'_1) \gamma^\mu \Phi^1_0 (p_1) \,
	{\bar \Phi}^1_0 (p_1) \gamma^\nu \Phi^1_0 (p'_1)
\nonumber\\
	&& \times
	{\bar\Phi}^2_0 (p'_2) \gamma^{\lambda} \Phi^2_0 (p_2) \,
	{\bar\Phi}^2_0 (p_2) \gamma^{\sigma} \Phi^2_0 (p'_2) \,
	|(2 \pi)^4 \delta (p_1{+}p_2{-}p'_1{-}p'_2)|^2 \,
	G^{(+)}_{\mu\lambda} (k) \, G^{(+)*}_{\nu\sigma} (k) \,.
\end{eqnarray}
where $k=p'_1-p_1$ is the momentum transfer and we have used
\begin{equation}
    G^{(-)}_{\mu\nu}(-k) = G^{(+)*}_{\mu\nu}(k) \,.
\end{equation}
The Debye screening effect due to the plasma is taken into
accounted by using the full two-point photon field Green's function
$G^{(+)} (k)$.
To illustrate this,
we consider a simple case which is a collision of heavy particles,
for example, protons, moving non-relativistically.
Letting $v_{\rm rel}$ be the relative velocity for defining the
cross section and using
\begin{equation}
    {\bar \Phi}^1_0 (p'_1) \gamma^\mu \Phi^1_0 (p_1)
	\approx \delta_{\mu 0} \, \Phi^{1\dagger}_0 (p'_1) \Phi^1_0 (p_1) \,,
	\quad {\bar \Phi}^2_0 (p'_2) \gamma^\mu \Phi^2_0 (p_2)
	\approx \delta_{\mu 0} \, \Phi^{2\dagger}_0(p'_2) \Phi^2_0 (p_2) \,,
\end{equation}
the cross section may be expressed as
\begin{eqnarray}
    \langle d \sigma \rangle &\approx& {q_1^2 q_2^2 \over v_{\rm rel}}
	\int_{\Omega_1} {d^3 p'_1 \over (2 \pi)^3}
	\int_{\Omega_2} {d^3 p'_2 \over (2 \pi)^3}
	(2 \pi)^4 \delta (p_1{+}p_2{-}p'_1{-}p'_2)
\nonumber\\
	&& \times |\Phi^{1\dagger}_0 (p'_1)\Phi^1_0 (p_1)|^2
        |\Phi^{2\dagger}_0 (p'_2)\Phi^2_0 (p_2)|^2 |G^{(+)}_{00} (k)|^2
\nonumber\\
	&\approx& {q_1^2 q_2^2 \over v_{\rm rel}}
	\int_{\Omega_1} {d^3 p'_1 \over (2 \pi)^3}
	\int_{\Omega_2} {d^3 p'_2 \over (2 \pi)^3}
	(2 \pi)^4 \delta (p_1 + p_2 - p'_1 - p'_2)
\nonumber\\
	&& \times
	|\Phi^{1\dagger}_0 (p'_1)\Phi^1_0 (p_1)|^2
        |\Phi^{2\dagger}_0 (p'_2)\Phi^2_0 (p_2)|^2
	{1 \over ({\bf k}^2{+}m_{\rm D}^2)^2}
\end{eqnarray}
where we have used to the approximated result:
\begin{equation}
    D_{00} (k_0=0, {\bf k}) \approx {1 \over {\bf k}^2 + m_{\rm D}^2}
\label{Debye}
\end{equation}
for ${\bf k}$ being small. Here $m_{\rm D}$ is the Debye mass.
When the momentum transfer is comparable with the inverse of the
Debye screening length, it probes Debye screening length in
the way above.

\subsection{Interaction between charged particle and the plasma via
soft photon}

In this subsection, we shall study the effect due to
the emission and absorption of the soft photons with energy and momentum
much less than the typical momentum transfer.
``Soft'' photons mean photons with energy and momentum much less that the
momentum transfer. For the hard photons interacting with the charged
particles, we can treat them perturbatively because of the smallness of
the QED coupling. However, since the soft photon causes
infrared divergences for individual diagram,
we need to sum up the leading infrared behavior\cite{Bloch&Nordsieck,Yennie}.
The process is always accompanied
by emission and absorption of a lot of soft photons. Treating it
perturbatively does not describe the process properly.
For the soft photon fields, we shall sum up the leading infrared
contribution to all order by using the Bloch-Nordsieck semi-classical
method.

The LSZ reduction formula enables us to extract the
scattering matrix element from Green's function.
Equivalently, we consider the large time behavior of the
Green's function and find the part having the correct
asymptotic wave function for the incoming and outgoing particles.
Mathematically,
\begin{eqnarray}
        && \prod_{f=1}^{n_f} \int dy_f \phi^{0\,*}_\h (p'_f, y_f)
        (\partial_{y_f}^2{+}m^2)
        \prod_{i=1}^{n_i} \int dx_i \phi^0_\h (p_i, x_i)
        (\partial_{x_i}^2{+}m^2)
	G[x_1,x_2,\cdots,x_{n_i},y_1,y_2,\cdots,y_{n_f}; A]
\nonumber\\
	&& \qquad = \prod_{f=1}^{n_f} \int dy_f
        \phi^*_\h [p_f, y_f; A] \prod_{i=1}^{n_i} \int dx_i
        \phi_\h [p_i, x_i; A]
	\Gamma [x_1,x_2,\cdots,x_{n_i},y_1,y_2,\cdots,y_{n_f}; A]
\end{eqnarray}
where $\phi_\h [p_i, x_i, A]$ and $\phi_\h [p_f, y_f, A]$ are the wave
functions of the charged particle moving in the classical
background $A$ field.
$\Gamma[x_1,x_2,\cdots,x_{n_i},y_1,y_2,\cdots,y_{n_f}; A]$
is the amputated Green's function for
the external charged particles or the
one-hard-momentum-particle-irreducible Green's function.
Since we focus on the contribution due to the soft photons
and the hard photon may be treated perturbatively, to the leading
order, we can neglect the hard photon contribution.
Equivalently, $A$ carries only very small energy and momentum
which are much less than the momentum transfer.
It is then valid to neglect soft photon field $A$ in
$\Gamma[x_1,x_2,\cdots,x_{n_i},y_1,y_2,\cdots,y_{n_f};A]$
as the leading infrared behavior is concerned
since $\Gamma$ describes a hard process\cite{Brown}.
Our major task now is to calculate $\phi_\h[p,x;A]$ for soft background
$A$ field by using classical current approximation.

For the reason of simplicity, let us first consider $\phi_\h[p,x;A]$ for
a scalar charged particle with charge $q$
satisfying the Klein-Gordon equation:
\begin{equation}
    (D_\mu D^\mu + m^2) \phi_\h [p,x; A] = 0 \,,
\end{equation}
where the covariant derivative $D_\mu = \partial_\mu - i q A_\mu$.
It is not hard to verify that, at the leading order, the
solution may be expressed as \cite{Brown}
\begin{equation}
    \phi_\h [p,x,A] = e^{i p x} \exp \Biggr \{
	i q \int_C dz_\mu A^\mu (z) \Biggr \}
\end{equation}
where $C$ is a straight line ending at $x_\mu$ with slope $p_\mu/m$
{\it i.e.} $C$ is the classical trajectory of the charged particle
moving with speed $v_\mu = p_\mu /m $.
It is generally true that the solution of a charged particle
with any spin moving in slowly varying background $A$ field
has the form:
\begin{equation}
    \phi_\h [p,x;A] = \phi^0_\h (p,x) \exp \Biggr \{
        i q \int_C dz_\mu A^\mu (z) \Biggr \}
\end{equation}
with $\phi^0_\h (p,x)$ being the free solution since
the effect due to spin is proportional to the field strength
$F_{\mu\nu}$ which is small\cite{Weinberg}.
It is convenient to define a classical current
\begin{equation}
    J_\mu (z) = q v_\mu \delta^{(3)} ({\bf z} - {\bf x}-
	{\bf v} (z_0 - x_0))
\end{equation}
so that we can write
\begin{equation}
    \phi_\h [p,x; A] = \phi^0_\h (p,x) \exp \Biggr \{
	i \int dz J_\mu (z) A^\mu (z) \Biggr \} \,.
\end{equation}

Now the soft photon problem can be simplified as the interaction
between classical current and the plasma.
The underlying reason for this is because the Compton wave length of the
charged particle is much smaller than the EM field's characteristic
wave length. Thus, at the scale of the wave length of the
EM field, the charged particle behaves like a classical particle.
This observation is originally due to
Bloch and Nordsieck\cite{Bloch&Nordsieck}.

Since $\Gamma[x_1,x_2,\cdots,x_{n_i},y_1,y_2,\cdots,y_{n_f};A]$
is the correlation function obtained by
integrating over the hard momentum, it certainly
looks local to the soft momentum. This means that
for the purpose of treating the soft photon, we
can regard it as proportional to a delta function
for the coordinates. Let us use $y \equiv y_1$ denote the
space-time point where the hard collision happens.
Each charged particle generates a current. The total current
is the sum of them. Let us denote the total current as $J^{(y)}$.
\begin{equation}
    J^{(y)}_\mu (z) = \theta (y_0{-}z_0)
	\sum_i q_i v^i_\mu \delta ({\bf z}{-}{\bf y}
	{-}{\bf v}^i (z_0{-}y_0))
	+ \theta (z_0{-}y_0) \sum_f q_f v^f_\mu
	\delta ({\bf z}{-}{\bf y}{-}{\bf v}^f (z_0{-}y_0) )
\label{J_def}
\end{equation}
and the Fourier transform $J_\mu (\omega, {\bf k})$ is simply
\begin{equation}
    J^{(y)}_\mu (\omega, {\bf k}) = i e^{i k y} \left (
	\sum_i q_i { p_\mu^i \over p^i{\cdot}k + i \epsilon}
	- \sum_f q_f {p_\mu^f \over p^f{\cdot}k - i \epsilon} \right ) \,,
\label{J_Fourier}
\end{equation}
where $k=(\omega,{\bf k})$.
The conservation of $J^{(y)}_\mu$ is guaranteed by $\sum_i q_i = \sum_f q_f$.
We now have,
\begin{eqnarray}
    &&\prod_{f=1}^{n_f} \int dy_f
        \phi^*_\h [p_f, y_f; A] \prod_{i=1}^{n_i} \int dx_i
        \phi_\h [p_i, x_i; A]
	\Gamma[x_1,x_2,\cdots,x_{n_i},y_1,y_2,\cdots,y_{n_f};A]
\nonumber\\
	&& \;\; \approx \prod_{f=1}^{n_f}\!\int\! dy_f
	\phi^{0*}_\h (p_f,y_f) \prod_{i=1}^{n_i}\!\int\! dx_i
        \phi^0_\h (p_i,x_i)
	\Gamma[x_1,x_2,\cdots,x_{n_i},y_1,y_2,\cdots,y_{n_f}; 0]
	e^{{i}\int dz J^{(y)}{\cdot}A} \,.
\end{eqnarray}
By translation invariance, we define ${\tilde \Gamma}$ as
\begin{eqnarray}
    &&\phi^{0*}_\h (p_1,y_1) \prod_{f=2}^{n_f} \int dy_f
        \phi^{0*}_\h (p_f,y_f) \prod_{i=1}^{n_i} \int dx_i
        \phi^0_\h (p_i,x_i)
        \Gamma[x_1,x_2,\cdots,x_{n_i},y_1,y_2,\cdots,y_{n_f}; 0]
\nonumber\\
	&& \qquad \equiv e^{i q y_1} {\tilde \Gamma}
	(p^i_1,p^i_2,\cdots,p^i_{n_i}; p^f_1,p^f_2,\cdots,p^f_{n_f})
\end{eqnarray}
where $q$ is the total four-momentum lost for the hard particles
due to the interaction with the plasma, i.e.
\begin{equation}
    q = \sum_{i=1}^{n_i} p^i - \sum_{f=1}^{n_f} p^f \,.
\end{equation}
The average rate including the correction due to the soft photon
is expressed as
\begin{equation}
    \rate \approx {1 \over T_\t} {1 \over V^{n_i}}
        \prod_{f=1}^{n_f} \int_{\Omega_f} {d^3 p_f \over (2 \pi)^3}
	|{\tilde \Gamma}
	(p^i_1,p^i_2,\cdots,p^i_{n_i}; p^f_1,p^f_2,\cdots,p^f_{n_f})|^2
	\int dy dy' e^{i q (y-y')} I (y,y')
\label{general_formula}
\end{equation}
where $I$ is
\begin{equation}
    I(y,y') = e^{J^{(y)} {\delta \over \delta J'}}
	e^{J^{(y')} {\delta \over \delta J'}}
	\left . e^{i W[J,J']} \right |_{J{=}0,J'{=}0}
	= e^{i W[J^{(y)},J^{(y')}]}\,. 
\end{equation}
with $J^{(y)}$ given by Eq.~(\ref{J_def}) and $J^{(y')}$ given by
\begin{equation}
    J^{(y')}_\mu (z) = J^{(y)}_\mu (z-y'+y) \,.
\end{equation}
To the leading order, we only need the quadratic term in the
expansion~(\ref{quadratic}) for $W[J,J']$.
$I(y,y')$ may be written as
\begin{eqnarray}
    I(y,y') &=& \exp \Biggr \{ - {i \over 2} \int \dk \int dt \int dt'
	\Big [J_\mu (t,{\bf k}) J_\nu (t',-{\bf k})
	{\cal G}_{\mu\nu} (t_>,t_<, {\bf k})
\nonumber\\
	&& \qquad  + J'_\mu (t,{\bf k}) J'_\nu (t',-{\bf k})
	{\cal G}_{\mu\nu} (t_<,t_>, {\bf k})
	- 2 J'_\mu (t,{\bf k}) J_\nu (t',-{\bf k})
	{\cal G}_{\mu\nu} (t',t, {\bf k}) \Big ]
	\Biggr \} \,.
\end{eqnarray}
Upon using the fact
\begin{equation}
    J_\mu'(t, {\bf k}) = J_\mu (t - (y_0 - y_0'), {\bf k}) \,
	e^{i {\bf k} \cdot ({\bf y} - {\bf y}')}
\end{equation}
and
\begin{equation}
    J_\mu' (\omega, {\bf k}) = J_\mu (\omega, {\bf k}) \,
	e^{i k (y'-y)} \,,
\end{equation}
we have
\begin{eqnarray}
    I (y,y') &=& \exp \Biggr \{ - i \int \dk \int dt \int dt'
	\left [J_\mu (t,{\bf k})
	- J'_\mu (t,{\bf k}) \right ] J_\nu (t',-{\bf k})
	{\cal G}_{\mu\nu} (t',t, {\bf k}) \Biggr \}
\nonumber\\
	&=& \exp \Biggr \{ - \int {d \omega \over 2 \pi} \int \dk
	J^*_\mu (\omega, {\bf k}) J_\nu (\omega, {\bf k})
	n(\omega) \rho_{\mu\nu} (\omega, {\bf k})
	\left (1 - e^{i k \cdot (y'-y)} \right ) \Biggr \} \,.
\label{I_result}
\end{eqnarray}

Since current $J$ is conserved, we can choose any gauge to work on.
In Landau gauge,
We shall use following decomposition to separate the longitudinal
part and the transverse part of the spectral density function:
\begin{equation}
    \rho_{\mu\nu} (\omega,{\bf k})
	= \rho_\t (\omega,{\bf k}) P^\t_{\mu\nu}
	+ \rho_\l (\omega,{\bf k}) P^\l_{\mu\nu}
\end{equation}
where
\begin{eqnarray}
    P^\t_{\mu\nu} &=& \delta_{\mu i} \left (\delta_{ij} -
	{{\bf k}_i {\bf k}_j \over {\bf k}^2} \right ) \delta_{j \nu}
\nonumber\\
    P^\l_{\mu\nu} &=& \left (\delta_{\mu 0} - {\omega {\bf k}_\mu
	\over k^2} \right ) \left (\delta_{\nu 0} - {\omega {\bf k}_\nu
        \over k^2} \right ) {k^2 \over {\bf k}^2} \,.
\end{eqnarray}
Defining ${\cal J}_{\t,\l}$
\begin{equation}
    {\cal J}_{\t,\l} \equiv J^*_\mu (\omega,{\bf k}) \,
	P_{\mu\nu}^{\t,\l} \, J_\nu (\omega, {\bf k})
\end{equation}
enables to express
\begin{equation}
    I (y,y') = \exp \Biggr \{ - \int {d \omega \over 2 \pi} \int \dk
	n(\omega) \left ({\cal J}_\t \, \rho_\t
	+ {\cal J}_\l \, \rho_\l \right )
	\left [1 - e^{i k \cdot (y'-y)} \right ] \Biggr \} \,,
\label{I_result2}
\end{equation}
Observing that $I(y,y')$ depends only on $y-y'$
in Eq.~(\ref{general_formula}), the space-time integrals
over $y$ and $y'$ gives the total volume of the space-time.
Therefore,
\begin{equation}
    \rate = {1 \over V^{n_i-1}} \prod_{i=1}^{n_f}
	\int_{\Omega_i} {d^3 p_f \over (2 \pi)^3}
	|{\tilde \Gamma} (p^i_1,p^i_2,\cdots,p^i_{n_i};
	p^f_1,p^f_2,\cdots,p^f_{n_f})|^2
        \int dy \, e^{- i q y} \, e^{- N (y)} \,.
\label{general_result}
\end{equation}
where we have defined $N(y)$
\begin{equation}
    N(y) \equiv \int {d \omega \over 2 \pi} \int \dk n(\omega)
	\left ({\cal J}_\t \, \rho_\t + {\cal J}_\l \, \rho_\l \right )
	\left (1 - e^{i k \cdot y} \right ) \,.
\label{N_def}
\end{equation}
In principle, one should be able to use the expression~(\ref{general_result})
to compute the effect on the hard process due to its interaction with the
plasma via soft photons. However, we shall not go to the detail
of the computation but only focus on two aspects, namely,
the infrared behavior and the absorption and emission of
soft plasmon.
Before generally discussing the infrared physics of
formula~(\ref{general_result}),
we like to study first the infrared behavior of
the damping of a charged particle moving
in QED plasma with hard momentum. The charged particle can
either move very fast or be very heavy so that its momentum
is hard.

\subsection{Damping of a charged particle}
\label{damping_subsec}

The question we ask is what is
the probability of detecting the charged particle moving with
the initial momentum at time $T_\t$, {\it i.e.} the probability
for finding the charged particle in the initial state.
We expect the answer may be expressed as an exponentially
decaying factor with the exponent proportional to the time $T_\t$. Thus
the half of the proportional factor in front of the time $T_\t$
in the exponent is the damping rate.
We also expect that the answer may depend on the momentum resolution
by which we judge whether the particle's momentum is the same as
the initial momentum since the soft photon emission
and absorption processes always appear.
We shall study the leading infrared contribution to the damping
of the charged particle due to its interaction with the plasma
via the exchange of soft photons.
Let $\Omega$, a small volume in momentum space, describe
the resolution.
By using the standard reduction technique, we can construct
the probability for finding the charged particle to have
momentum ${\bf p}'$ lying in the phase space volume $\Omega$
centered at ${\bf p}$. Keeping the leading infrared terms,
it may be expressed as
\begin{equation}
    P = \int_{\Omega} {d^3 {\bf p}' \over (2 \pi)^3}
        \int d {\bf y} \, e^{i ({\bf p}-{\bf p}') \cdot {\bf y}}
	\, e^{- N(0,{\bf y})}
\end{equation}
where
\begin{equation}
    N (0,{\bf y}) = \int {d \omega \over 2 \pi} \int \dk n(\omega)
        J^*_\mu (\omega, {\bf k}) J_\nu (\omega, {\bf k})
        \rho_{\mu\nu} (\omega, {\bf k})
        \left (1 - e^{i {\bf k} \cdot {\bf y}} \right )
\end{equation}
with the current
\begin{equation}
    J_\mu (\omega, {\bf k})
        = \int_0^{T_{\rm t}} dt \int d{\bf x} \,
	e^{i \omega t - i {\bf k} {\bf x}}
	\, q \, v_\mu \delta ({\bf x} - {\bf v} t)
        = q \, v_\mu {e^{i (\omega - {\bf v} \cdot {\bf  k}) T_{\rm t}}
        -1 \over i (\omega - {\bf v} \cdot {\bf  k})}
\end{equation}
Noting that 
\begin{equation}
    J^*_\mu (\omega, {\bf k}) J_\nu (\omega, {\bf k})
        \to q^2 v_\mu v_\nu T_\t \, (2 \pi) \,
        \delta (\omega - {\bf v} \cdot {\bf  k}) \quad
        {\rm as} \;\; T_\t \to \infty
\end{equation}
we can perform the integral over $\omega$ and get
\begin{equation}
    N(0,{\bf y}) = q^2 T_\t \int \dk n({\bf k} \cdot {\bf v}) v_\mu v_\nu
        \rho_{\mu\nu} ({\bf k} \cdot {\bf v}, {\bf k})
        \left (1 - e^{i {\bf k} \cdot {\bf y}} \right ) \,.
\end{equation}
We shall look at the leading infrared behavior. The spectral
density $\rho_{\mu\nu}$ is dominated by its transverse part
$\rho_t$ at the infrared region. As $|{\bf k}|$ becomes small,
the leading infrared behavior for $\rho_t$ is\cite{Pisarski}
\begin{equation}
    {\rho_t (\omega, {\bf k}) \over \omega}
	\to {1 \over {\bf k}^2} (2 \pi) \delta (\omega) \,.
\label{rhot_asy}
\end{equation}
Using this, we can complete the integral over the angular part for
${\bf k}$ integral as
\begin{eqnarray}
    N(0,{\bf y}) &=& q^2 T \, T_\t
        \int_0^{\infty} e^{-k/ \Lambda} {dk \over 4 \pi^2 k} |{\bf v}|
        \int_0^{2 \pi} d \phi
        \left (1 -  e^{i k |{\bf y}| \sin\theta\cos\phi} \right )
\nonumber\\
	&=& q^2 \, T \, T_\t
	\int_0^{2 \pi} d \phi {1 \over 4 \pi^2}
	|{\bf v}| \ln (1 - i \Lambda |{\bf y}| \sin\theta\cos\phi )
\nonumber\\
	&\approx& {q^2 \over 2 \pi} T \, T_\t \,
	|{\bf v}| \ln (\Lambda |{\bf y}| \sin\theta)
	= {q^2 \over 2 \pi} T \, T_\t \,
        |{\bf v}| \ln (\Lambda |{\bf y}_\t|) \,,
\label{Ndamping}
\end{eqnarray}
where an ultraviolet cutoff $\Lambda$ for the $k$ integral
has been introduced, $\theta$ is the angle between ${\bf y}$
and the velocity of the charged particle ${\bf v}$, and
${\bf y}_\t \equiv {\bf y} - ({\bf y}{\cdot}{\hat {\bf v}}){\hat {\bf v}}$.
Therefore,
\begin{equation}
    P \sim \left ({\Delta p_\t \over \Lambda} \right )^{ {q^2
	 \over 2 \pi} T |{\bf v}| T_\t}
	= \exp \left \{- {q^2 \over 2 \pi} T |{\bf v}|
	\ln \left ({\Lambda \over \Delta p_\t} \right ) T_\t \right \}  \,,
\end{equation}
where $\Delta p_\t$ is the transverse size of $\Omega$ with respect
to the velocity ${\bf v}$ of the charged particle.
At the leading order, the damping rate is
\begin{equation}
    \Gamma_{\rm damping} = {q^2 \over 4 \pi} T |{\bf v}|
        \ln \left ({\Lambda \over \Delta p_\t} \right ) \,.
\label{damping_rate_result}
\end{equation}
The damping rate depends on the momentum resolution $\Delta p_\t$
logarithmically due to the exchange of soft magnetic photon
since the magnetic force is not screened in QED plasma
and is transverse to ${\bf v}$.
This leading piece of the damping rate vanishes if ${\bf v}={\bf 0}$
which is the case where the charged particle does not feel the
magnetic force. If we take the limit $\Delta p_\t \to 0$, the damping
rate becomes infinite. This means that the particle experiences
the long range magnetic force and can not move with a definite
momentum. There has been discussions about the infrared problem
for the damping rate of charged particles moving in QED plasma
in the literature\cite{Lebedev&Smilga&Baier}. This infrared problem can be
resolved by considering only a physically measurable rate which
introduces the momentum resolution $\Delta p_\t$ to cut off the
infrared divergence.

To avoid confusion, we note that only the infrared behavior of
the damping rate has been examined. Equation~(\ref{damping_rate_result})
is valid only when the cutoff $\Lambda$ being larger than $\Delta p_\t$.
Cutoff $\Lambda$ is the scale below which the asymptotic behavior
\ref{rhot_asy} can be used. At high temperature, $\Lambda \sim eT$
while at low temperature $\Lambda$ is suppressed by factor
$e^{-\beta m_\e}$ with $m_\e$ being the electron mass.
Therefore, at temperature $T \ll m_\e$, this infrared cutoff
becomes almost zero. Our resolution $\Delta p_\t$ will not
be able to see this scale. Thus, the ``leading infrared
behavior'' is essentially negligible. In view of Eq.~(\ref{Ndamping}),
this corresponds to the case $\Lambda |{\bf y}| \ll 1$ so that
$N(0,{\bf y})$ is small. These comments apply to discussions in
next subsection as well.

\subsection{General discussion of the infrared behavior}
The cancellation of infrared divergences have been studied for
the case of ideal thermal photon gas in reference\cite{Weldon1}.
This is the case where the temperature is low and the chemical
potential for the electron is small so that the photon gas
is almost ideal.
Here we shall provide a general discussion of the infrared
physics. Using the result~(\ref{J_Fourier}),
\begin{equation}
    {\cal J}^{\t,\l}
	= \left (\sum_i {q_i \, p_{i\mu} \over p_i{\cdot}k{+}i \epsilon}
	- \sum_f {q_f \, p_{f\mu} \over p_f{\cdot}k{-}i \epsilon} \right )
	\left (\sum_i {q_i \, p_{i\nu} \over p_i{\cdot}k{-}i \epsilon}
	- \sum_f {q_f \, p_{f\nu} \over p_f{\cdot}k{+}i \epsilon} \right )
	P^{\t,\l}_{\mu\nu} \,.
\end{equation}
Putting this into the definition~(\ref{N_def}) for integral $N(y)$,
it is not hard to find that there are singularities
in the integral over $\omega$ or ${\bf k}$ corresponding to pinching
two poles together as $\epsilon \to 0$.
The physical meaning of these singularities which come from putting
the charged particle's propagator on-shell is that the charged particle
suffers the damping as discussed in the previous subsection.
Therefore, their contributions to $N(y)$ can be expressed similarly
as the result shown in Eq.~(\ref{Ndamping}).
We make following replacement to subtract out the damping
effect of the single particle:
\begin{equation}
    J^*_\mu (k) J_\nu (k) \to
        J^*_\mu (k) J_\nu (k) - \sum_i {q_i^2 \; p_{i\,\mu} p_{i\,\nu}
	\over (p_i \cdot k + i \epsilon) (p_i \cdot k - i \epsilon)}
        - \sum_f {q_f^2 \; p_{f\,\mu} p_{f\,\nu}
	\over (p_f \cdot k - i \epsilon) (p_f \cdot k + i \epsilon)} \,.
\end{equation}
Because of Debye screening, the exchange of the
magnetic photons dominate at the infrared region for
the $\omega,{\bf k}$ integrals..
By the sum rule\cite{Pisarski}
\begin{equation}
     \int {d \omega \over 2 \pi} \,
	{\rho_\t (\omega, {\bf k}) \over \omega}
	\sim {1 \over {\bf k}^2}
\end{equation}
and the behavior
\begin{equation}
    {\rho_\t (\omega, {\bf k}) \over \omega}
	\to {1 \over {\bf k}^2} (2 \pi) \delta (\omega)
\end{equation}
as ${\bf k} \to 0$,
we have the leading infrared contribution to $N(y)$
\begin{equation}
    N(y)_{\rm infra} \approx T \int \dk {1 \over {\bf k}^2} \left (
	{\bf J}^* (0,{\bf k}){\cdot}{\bf J} (0,{\bf k})
	- \sum_i \left |{q_i \over
	{\hat {\bf v}}_i \cdot {\bf k}{+}i \epsilon} \right |^2
	- \sum_f \left |{q_f \over {\hat {\bf v}}_f \cdot {\bf k}
	{-} i \epsilon} \right |^2 \right )
	(1{-}e^{i {\bf k} \cdot {\bf y}})
\end{equation}
besides the contribution from the damping of the individual particles.
The ${\bf k}$ integral in expression above
may be written as a one-parameter integral as shown in the appendix.
Using the conservation of the current so as to
replace $P_{\mu\nu}^\t$ by $\delta_{ij}$, apart from the contribution
due to the damping effect,
\begin{eqnarray}
    N(y)_{\rm infra} &\approx& T |{\bf y}| \left [
	\sum_{i=1}^{n_i}\sum_{i'=i{+}1}^{n_i} q_i  q_{i'}
	I({\hat {\bf y}},{\hat {\bf v}}_i,{-}{\hat {\bf v}}_{i'}) \right .
	+ \sum_{f=1}^{n_f}\sum_{f'=f{+}1}^{n_f} q_f  q_{f'}
	I({\hat {\bf y}},{-} {\hat {\bf v}}_f,{\hat {\bf v}}_{f'})
\nonumber\\
	&& \qquad \qquad \left . + \sum_{i=1}^{n_i} \sum_{f=1}^{n_f}
	q_i q_f I({\hat {\bf y}},{\hat {\bf v}}_i,{\hat {\bf v}}_f)
	\right ] \,,
\end{eqnarray}
where $I({\hat {\bf y}}, {\bf v}, {\bf v}')$ is defined and simplified
in the appendix. Since $I({\hat {\bf y}}, {\bf v}, {\bf v}')$ contains
no infrared divergence, $N(y)_{\rm infra}$ is finite.
The unscreened static magnetic force in QED plasma
causes no infrared divergences.

\subsection{long wave plasma oscillation}
Besides to the leading infrared effect due to the unscreened
static magnetic force, there are also effects due to 
the excitations of the long wave plasma oscillation.
To simplify the discussion, let us isolate the contribution of
the plasmon by considering
\begin{equation}
    \rho_{\t,\l} = (2 \pi) \delta (\omega^2 - \omega_{\t,\l} ({\bf k})^2)
\end{equation}
with $\omega = \omega_{\t,\l} ({\bf k})$ being the dispersion
relation for the transverse and longitudinal plamons respectively.
Putting this into the expression~(\ref{N_def}) gives
\begin{eqnarray}
    N(y)_{\rm pl} &=& \int \dk \left [{{\cal J}_\t \over 2 \omgtk}
	\Big [(1 - e^{i k_\t y}) + 2 n(\omgtk)
	(1 - \cos k_\t y) \Big ] \right .
\nonumber\\
	&& + \left. {{\cal J}_\l \over 2 \omglk}
	\Big [(1 - e^{i k_\l y}) + 2 n(\omglk)
	(1 - \cos k_\l y) \Big ] \right ] \,,
\label{plasmon}
\end{eqnarray}
where $k_{\t,\l} = (\omega_{\t,\l}({\bf k}), {\bf k})$.
For the integrals in Eq.~(\ref{plasmon}), the plasma mass
$m_{\rm pl} \equiv \omega_{\t,\l}({\bf k}={\bf 0})$ cuts
off the integral at the infrared region.
At very high temperature $m_{\rm pl} \sim eT$ or
at low temperature with sufficiently large chemical potential,
$m_{\rm pl} \gg e^2 T$, $N(y)$ is of order
$e^2 T / m_{\rm pl} \ll 1$, thus perturbation theory works well
since we do not need to sum over the series to form the small exponential
factor.

At low temperature and small chemical potential,
$m_{\rm pl}$ may be much smaller than $e^2 T$.
There are two cases that we would like to discuss.

1) For the case where the exchange of energy between the
charged particles and the plasma is much smaller or
the same order of magnitude of $m_{\rm pl}$, we then
need to compute integral~(\ref{plasmon}) by using, for example,
the one-loop spectral density function. The calculation is rather
involved. We leave it to another day.

2) For the case where the exchange of energy between the
charged particles and the plasma is much larger than
$m_{\rm pl}$, we can set $m_{\rm pl}$ to be zero and simply
use the dispersion relation $\omega_{\t,\l} ({\bf k}) = |{\bf k}|$.
For this case, the contribution to $N(y)$ from the plasmon dominates
since the photons are nearly ideal.
Therefore,
\begin{equation}
    N(y) \approx N(y)_{\rm pl} = \int \dk {1 \over 2 |{\bf k}|}
	\left | J_\mu (|{\bf k}|,{\bf k}) \right |^2
	\left [ (1 - e^{i k y}) + 2 n(|{\bf k}|)
	(1 - \cos k y) \right ] \,.
\label{massless}
\end{equation}
Writing
\begin{equation}
    \left | J_\mu (|{\bf k}|,{\bf k}) \right |^2
	= {1 \over {\bf k}^2} 
	\left |\sum_{i=1}^{n_i} {q_i \, p_{i\mu} \over p_i{\cdot}u}
	- \sum_{f=1}^{n_f} {q_f \, p_{f\mu} \over p_f{\cdot}u} \right |^2
\end{equation}
with four-vector $u \equiv (1,{\hat {\bf k}})$, we have
\begin{equation}
    N(y) \approx {1 \over 8 \pi^3} \int d{\hat {\bf k}}
	\left |\sum_{i=1}^{n_i} {q_i \, p_{i\mu} \over p_i{\cdot}u}
	- \sum_{f=1}^{n_f} {q_f \, p_{f\mu} \over p_f{\cdot}u} \right |^2
	\int_0^\infty {d \omega \over 2 \omega}
        \left [ (1 - e^{i \omega u {\cdot} y})
	+ {2 \over e^{\beta \omega}{-}1}
        (1{-}\cos \omega u{\cdot}y) \right ] \,.
\label{massless2}
\end{equation}
Expression~(\ref{massless2}) above contains an ultraviolet divergence which
should be cut off at the hard momentum scale $\Lambda$. Instead of using a
sharp cut off, we use an exponential factor $e^{-\omega/\Lambda}$
to suppress the contribution from the region with $\omega$
bigger than $\Lambda$.
Using the result in the appendix for the $\omega$ integral,
\begin{equation}
    N(y) \approx {1 \over 16 \pi^3} \int d{\hat {\bf k}}
	\left |\sum_{i=1}^{n_i} {q_i \, p_{i\mu} \over p_i{\cdot}u}
	- \sum_{f=1}^{n_f} {q_f \, p_{f\mu} \over p_f{\cdot}u} \right |^2
        \left [ \ln \left (1{+}i \Lambda u {\cdot} y \right )
	+ \ln {\sinh \pi T u{\cdot}y \over \pi T u{\cdot}y} \right ] \,.
\label{massless_exp}
\end{equation}
A full analysis of expression~(\ref{massless_exp}) is still formidable.
We shall consider a special case where the process involves at least
one heavy particle moving non-relativistically so that it can absorb
the extra momentum without affect much the conservation of energy.
Thus, we effectively has only the energy conservation
without worrying about where to dump the extra momentum.
For this case, we can effectively replace ${\bf y}$ by ${\bf 0}$
and only left with the time component.
It is convenient to consider the differential
rate with the energy exchange with the plasmons in plasma to be
in the interval $(E,E+dE)$:
\begin{eqnarray}
    d\rate &\propto& \int dt e^{-i E t} \exp \Biggr \{
	- A \left [ \ln (1 + i \Lambda t)
	+ \ln {\sinh \pi T t \over \pi T t} \right ]
	\Biggr \} {dE \over 2 \pi}
\nonumber\\
	&=& \int dt e^{-i E t}
	(1 + i \Lambda t)^{- A} \left ({\sinh \pi T t \over \pi T t}
	\right )^{-A} {dE \over 2 \pi}
\label{massless_diffrate}
\end{eqnarray}
where $A$ is defined by
\begin{equation}
    A \equiv {1 \over 16 \pi^3} \int d {\hat {\bf k}}
	\left (\sum_{i=1}^{n_i} q_i {p_{i\mu} \over p_i{\cdot}u}
	- \sum_{f=1}^{n_f} q_f {p_{f\mu} \over p_f{\cdot}u}
	\right )^2 \,.
\end{equation}

We can then reproduce two previous results\cite{Yennie,Weldon2}.
For the case $T \ll E$, it is expected that the result
is approximately the same as the zero temperature case.
For the $t$ integral in expression~(\ref{massless_diffrate}),
typical $t$ is of scale $1/E$ and thus $T t \ll 1$.
Noting $ \sinh x \approx x $ for small $x$, we have
\begin{equation}
    d \rate \propto \int dt e^{-i E t}
        (1 + i \Lambda t)^{- A} \, {dE \over 2 \pi}
	\sim dE \, \theta(E) \, {A \over E} \,
	\left ({E \over \Lambda} \right )^A \,.
\end{equation} 
Therefore, the transition rate with soft photon carrying away
energy no more than $\Delta E$ with $\Delta E \gg T$ is
\begin{equation}
    \rate_{\Delta E} \, \propto \int_0^{\Delta E} dE \,
	{A \over E} \, \left ({E \over \Lambda} \right )^A
	= \left ({\Delta E \over \Lambda} \right )^A \,.
\end{equation}
This is the zero temperature result\cite{Yennie}.

For the case where the energy exchange
between plasma and charged particles are much smaller than the
temperature, $T \gg E$. Now for typical $t$ in the integral
in Eq.~(\ref{massless_diffrate}), $Tt \gg 1$. Using
$\sinh x \sim e^x$ for large $x$,
\begin{eqnarray}
    d \rate &\propto& {dE \over 2 \pi}
        \int dt e^{-i E t} e^{- \pi A T |t|}
	\left ({T \over \Lambda} \right )^A
\nonumber\\
	&=& {dE \over 2 \pi} \left ({T \over \Lambda} \right )^A
	\left [{1 \over \pi A T{+}i E } +
	{1 \over \pi A T{-}i E } \right ]
\nonumber\\
	&\approx& {dE \over 2 \pi} {2 \pi A T \over E^2 + (\pi A T)^2} \,.
\end{eqnarray}
The agrees with previous result\cite{Weldon2}.

\section{Conclusions}
A general formula for computing the transition rate for
hard process happening in a soft thermal background is
derived. In particular, we use it to study hard process
involving charged particle inside QED plasma.
Both the Debye screening and the damping of the hard
charged particles are incorporated in our formula.
We show that the transition rate is free of infrared divergence
even though the magnetic force is not screened for QED plasma.
Soft plasmon emission and absorption are analyzed.
Our formalism can also be applied to the case where some or all of
the hard momentum particles are replaced by particles which
interact weakly with the plasma so that after the formation of
the plasma, it does not have enough time to
have thermal contact with the degree of freedom corresponding
to these particles.

%%%%%%%%%%%%%%%%%%%%%%%%%%%%%%%%%%%%%%%%%%%%%%%%%%%%%%%%%%%%%%%%%%%%%%%%%%

\acknowledgements

I would like to thank L.~S.~Brown for bringing me relevant references.
I am also grateful to T.~Clark and S.~Love for discussions.
This work was supported by the U.S. Department of Energy,
grant DE-AC02-76ER01428 (Task B).

%%%%%%%%%%%%%%%%%%%%%%%%%%%%%%%%%%%%%%%%%%%%%%%%%%%%%%%%%%%%%%%%%%%%%%%%%%
%%%%%%%%%%%%%%%%%%%%%%%%%%%%%%%%%%%%%%%%%%%%%%%%%%%%%%%%%%%%%%%%%%%%%%%%%%

\newpage

\appendix

\section{Computation of integrals}

We first evaluate following integral:
\begin{equation}
    I({\hat {\bf y}}, {\bf v}, {\bf v}')
	\equiv {1 \over |{\bf y}|}
	\int \dk {1 \over {\bf v} \cdot {\bf k} + i\epsilon}
	{1 \over {\bf v}' \cdot {\bf k} + i\epsilon}
	{1 \over {\bf k}^2}
	\left [2 - e^{i {\bf k} \cdot {\bf y}}
	- e^{-i {\bf k} \cdot {\bf y}} \right ] \,.
\end{equation}
Exponentiating the denominators produces
\begin{equation}
    I({\hat {\bf y}}, {\bf v}, {\bf v}')
	= - {1 \over |{\bf y}|}
	\int_0^\infty dt \int_0^\infty dt'
	\int \dk {1 \over {\bf k}^2}
	e^{i {\bf k} \cdot ({\bf v} t + {\bf v}' t')}
	\left [2 - e^{i {\bf k} \cdot {\bf y}}
        - e^{-i {\bf k} \cdot {\bf y}} \right ] \,.
\end{equation}
Completing the ${\bf k}$ integral gives
\begin{equation}
    I({\bf y}, {\bf v}, {\bf v}')
        = - {1 \over |{\bf y}|}
	\int_0^\infty dt \int_0^\infty dt'
	\left [{2 \over |{\bf v} t + {\bf v}' t'|}
	- {1 \over |{\bf v} t + {\bf v}' t'+{\bf y}|}
	- {1 \over |{\bf v} t + {\bf v}' t'-{\bf y}|}
	\right ] \,.
\end{equation}
We can insert another integral over $s$ and change the variables
$t=s \alpha, t'=s \beta$ to get
\begin{eqnarray}
    I({\hat {\bf y}}, {\bf v}, {\bf v}')
	&=& - \int_0^\infty {ds \over |{\bf y}|}
	\delta (s{-}t{-}t') \int_0^\infty dt \int_0^\infty dt'
        \left [{2 \over |{\bf v} t{+}{\bf v}' t'|}
        - {1 \over |{\bf v} t{+}{\bf v}' t'{+}{\bf y}|}
        - {1 \over |{\bf v} t{+}{\bf v}' t'{-}{\bf y}|}
        \right ]
\nonumber\\
	&=&{-}\int_0^\infty \! {ds \over |{\bf y}|} \int_0^1 d \alpha
	\left [{2 \over |\alpha {\bf v}{+}(1{-}\alpha) {\bf v}'|}
	{-}{1 \over |\alpha {\bf v}{+}(1{-}\alpha) {\bf v}'{+}{\bf y}/s |}
	{-}{1 \over |\alpha {\bf v}{+}(1{-}\alpha) {\bf v}'{-}{\bf y}/s |}
	\right ]
\nonumber\\
	&=& \int_0^1 d \alpha {1 \over \left [
	\alpha {\bf v}{+}(1{-}\alpha) {\bf v}' \right ]^2}
	\int_0^\infty d s \left [
	{s \over \sqrt{1{-}2 s \cos \theta(\alpha){+}s^2}}
	{+} {s \over \sqrt{1{+}2 s \cos \theta(\alpha){+}s^2}}
	{-}2 \right ]
\nonumber\\
	&=& \int_0^1 d \alpha {1 \over \left [
        \alpha {\bf v}{+}(1{-}\alpha) {\bf v}' \right ]^2}
	\left [2 + \cos\theta(\alpha) \ln
	{1 + \cos\theta(\alpha) \over 1 - \cos\theta(\alpha)}
	\right ] \,,
\end{eqnarray}
where $\theta(\alpha)$ is the angle between vectors
$\alpha {\bf v}{+}(1{-}\alpha) {\bf v}'$ and ${\bf y}$.

We now evaluate integral
\begin{equation}
    I(\eta) \equiv \int_0^\infty
	{dx \over x} {2 \over e^x -1} (1 - \cos \eta x) \,.
\end{equation}
Expanding the denominator gives
\begin{equation}
    I(\eta) = \sum_{n=1}^\infty \int_0^\infty
	{dx \over x} \left [2 e^{-nx} - \left (
	e^{-nx+i\eta x} - e^{-nx - i\eta x} \right ) \right ]
	= \sum_{n=1}^\infty \ln \left (1 + {\eta^2 \over n^2} \right )
	= \ln {\sinh \pi \eta \over \pi \eta} \,,
\end{equation}
where we have used
\begin{equation}
    \int_0^\infty {dx \over x} (e^{-a x} - e^{-b x})
	= \ln {a \over b}
\end{equation}
and the infinite product
\begin{equation}
    \sinh x = x \prod_{n=1}^\infty
	\left (1{+}{x^2 \over \pi^2 n^2} \right ) \,.
\end{equation}

\end{document}